\documentclass[article]{has40}
\usepackage{graphicx}
\usepackage{amssymb}
\tabletypesize{\normalsize}  % AMcW

\shortauthors{Kolenberg}
\shorttitle{RR Lyrae Studies with Kepler}

\begin{document}
\large    %AMcW  The conference proceedings will employ large size print
\pagenumbering{arabic}
\setcounter{page}{74}

\title{RR Lyrae Studies with Kepler}

\author{{\noindent Katrien Kolenberg{$^{\rm 1,2}$}}\\
\\
{\it (1) Harvard-Smithsonian Center for Astrophysics, 60 Garden Street, Cambridge MA 02138, USA, 
(2) Instituut voor Sterrenkunde, K.U. Leuven, Celestijnenlaan 200D, 3001 Heverlee, Belgium} 
}

\email{(1) kkolenberg@cfa.harvard.edu}

\begin{abstract}
The spectacular data delivered by NASA's {\it Kepler} mission have not only boosted the discovery of planets orbiting other stars, but they have opened a window on the inner workings of the stars themselves. For the study of the RR Lyrae stars, Kepler has led to a breakthrough. To date, over 50 RR Lyrae stars are known in the Kepler field. They are studied within the RR Lyrae/Cepheid working group of the Kepler Asteroseismic Science Consortium (KASC).
This paper highlights some of the most interesting results on RR Lyrae stars obtained through Kepler so far.
\end{abstract}

\section{Kepler and the RR Lyrae stars}
The {\it Kepler} mission was launched in March 2009, with the primary goal to find Earth-like planets around other stars.  In searching for those, it delivers data of unprecedented precision for variable star studies, and asteroseismology in particular.  Most of the pulsating stars observed with the Kepler mission have been studied within the framework of the Kepler Asteroseismic Science Consortium (KASC)\footnote{http://astro.phys.au.dk/KASC/}, in which working groups are focusing on specific types of pulsating stars. 
%(http://astro.phys.au.dk/KASC/).  
The RR Lyrae working group has been doing time-series analysis of the Kepler RR Lyrae data, ground-based follow-up observations, and theoretical modeling inspired by the obtained observational results. Since Kepler has entered its extended mission time, the RR Lyrae and Cepheid working groups have merged to combine expertise on similar types of pulsators (and because only one Cepheid is known in the Kepler field).

At the end of Kepler's nominal mission in November 2012, 41 bona-fide RR Lyrae stars were known in the Kepler field and put on the target list for the extended mission (most in long cadence, three in short cadence).
Since then, several new RR Lyrae stars of different types (non-modulated RRab stars, Blazhko RRab stars, and RRc stars) have been serendipitously found in the Kepler field, many of them thanks to the citizen science project "PlanetHunters"\footnote{ www.planethunters.org}. Some were found in the Kepler super-apertures (a number of regions in the field observed through large, custom apertures; they typically cover open clusters and background regions devoid of bright stars). It is likely that with time more RR Lyrae stars will be found in the Kepler field. 

Kepler data come in two varieties: the long-cadence data yielding one data point every 30 minutes, and the short, 1-minute cadence data. Whereas the long-cadence data are sufficient to cover the light curve quite well, features such as the hump (a change in slope often occurring on the rising branch towards maximum light around pulsation phase 0.95) and the bump (occurring on the descending branch before minimum light around pulsation phase 0.7) can only be studied with short-cadence data.  We made sure to obtain at least one quarter (three months) of short cadence data for each RR Lyrae star known in the Kepler field during the nominal mission.

\section{Some unexpected results}

The unprecedented precision of the Kepler data and their nearly-uninterrupted coverage over several years, have led to multiple new findings. Several of those were reported on in previous reviews, see, e.g., Kolenberg (2011). 

Table 1 lists papers that were published based on or inspired by the Kepler data on RR Lyrae stars, either through direct analysis of the Kepler light curves (marked "A" for analysis), sometimes in combination with modeling of the observations or pure modeling papers (marked "M" for modeling), or in combination with ground-based follow-up data of the stars (marked "G" for ground-based data). The number of Kepler RR Lyrae stars studied in each paper is also added. Several papers were inspired by some of the findings in the unprecedentedly precise Kepler data (marked "S" for spin-off).
This list is not exhaustive (apologies for unintended omissions) and will grow, as the Kepler database continues to be mined. Below I highlight a few of the - partly unexpected - results, obtained with the Kepler data.

\subsection{Highlight 1: Flavors of Variability - Remarkably Stable Stars and The Blazhko Zoo}

The non-modulated stars in the Kepler field are remarkably stable below mmag level.  Nemec et al. (2011) analyzed these non-Blazhko RR Lyrae stars and derived photometric [Fe/H] values through Fourier correlations. 

Close to 50\% of the RRab stars in the Kepler field show Blazhko modulation, and a large variety of Blazhko behavior is observed (in period range, multiple modulation periods, non-repeating modulation, see Benk\H{o} et al. 2011).  In addition, additional frequencies are found, both at the expected postions of higher-order radial overtones, and at other positions. Two stars are found to pulsate in their fundamental and second overtone modes. 

\subsection{Highlight 2: RR Lyr itself}

The fact that RR Lyr, the prototype and arguably the best-studied star of the class, is located in the Kepler field, is a very fortunate coincidence.  Thanks to a custom aperture  (described in Kolenberg et al 2011) we can obtain accurate photometry of this star, even though it was initially considered to be too bright to be observed with Kepler. The star was observed in Q1 and Q2 (May 13 - September 16, 2009) in long cadence and from Q5 onwards (March 20, 2010) in short cadence. 
It turns out that the prototype of the class is a Blazhko star showing features that could only be detected with precise satellite data:  period doubling (see Kolenberg et al. 2010, Szabo et al. 2010) and more complex frequency patterns (Molnar et al. 2012), non-repeating Blazhko cycles (Kolenberg et al. 2011) as well as evidence of the presence of the first overtone (Moln\'ar et al. 2012).  
%Plachy et al. (2013) and Smolec et al. (2013) show that low-dimensional chaos may play a role in RR Lyr's not-clockwork-regular pulsation patterns.
Stellingwerf et al. (these proceedings) present their interesting findings on the short-cadence data of RR Lyr. For the first time we see evidence of a cyclic variation of the Blazhko period, and a direct recording of the star's four-year cycle, first mentioned by Detre \& Szeidl (1973).  

\begin{figure*}
\begin{center}
\includegraphics[scale=0.5, angle=-90]{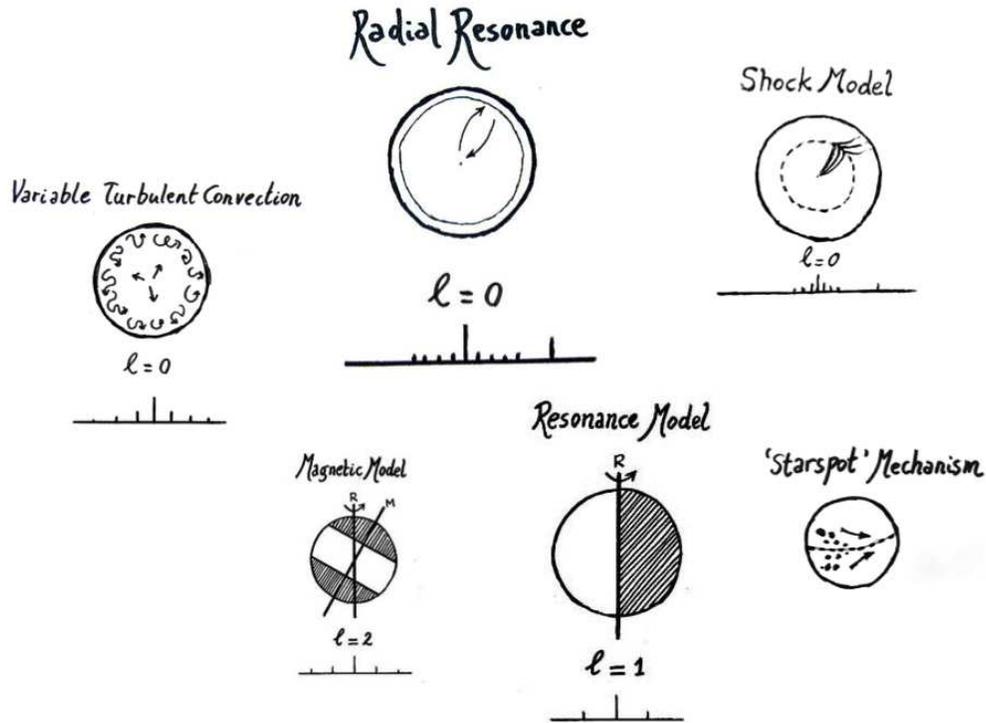}
\vskip0pt
\caption{Schematic representation of several recently (re)proposed models for the Blazhko.}
\end{center}
\end{figure*}

\section{Highlight 3: Kepler and the models for the Blazhko effect}

The findings based on the Kepler data have led to a restructuring in the hierarchy for the models for the Blazhko effect (see Kolenberg 2011). 
First of all, the observation of period doubling in several Kepler Blazhko stars and investigations towards their origin led to the radial resonance model for the Blazhko effect. In this model, the 9:2 resonance responsible for period doubling can also lead to modulation or the Blazhko effect (Buchler \& Koll\'ath 2011). To date, this model is the most commonly quoted for explaining the Blazhko effect in RR Lyrae stars. 

The scenario proposed by Stothers (2006, 2010), based on variable turbulent convection, was put to the test by Smolec et al. (2010) and Moln\'ar et al. (2010). 
A new model involving shock waves was recently proposed by Gillet (2013). 
The contribution by Stellingwerf, Nemec \& Moskalik (this proceedings) mentions an alternative that explains a variable Blazhko period through differential rotation ("starspot" mechanism, see also Detre \& Szeidl 1973). The underlying mechanism for the Blazhko effect may be described by elements from several models.  A schematic representation of the models recently mentioned is shown in Figure\,1.

\pagebreak
\section{Highlight 4: RRc stars}

One of the most puzzling discoveries with Kepler was made based on the Kepler RRc star data.  For each of the four RRc stars analyzed in detail so far, an additional frequency is detected, besides the main frequency corresponding to the radial first overtone and its harmonics.  This additional period always has a ratio of about 0.61 to the main radial mode (first overtone) period.  Moreover, we also see period doubling for this additional frequency. The period doubling seen here is notably different from the one observed in RRab Blazhko stars, where the main radial mode shows period doubling. 
The observed period ratio does not correspond to any of the ratios known for radial modes in RR Lyrae stars and models, hence it is suggested that this additional frequency may be related to a nonradial mode.  However, it is striking that we see similar ratios in each of the studied stars. This ratio has also been observed in Cepheids (see Moskalik et al. (2012) and references therein).

In the meantime a few more RRc stars are known in the Kepler field and their light curves remain to be analyzed.

\section{Highlight 5: Ground-based data}

For all of the known stars in the Kepler field during the nominal mission the KASC RR Lyrae working group has obtained complementary data obtained with ground-based telescopes.  These include multicolor photometric data (e.g., Jeon et al. 2013), and high-resolution spectroscopic data (Nemec et al. 2013).   They allow us to characterize the stars in the Kepler field much better. Nemec et al. (2013) derived spectroscopic [Fe/H] values for 34 stars ranging from very metal-poor ([Fe/H] $\simeq$ -2.5) to nearly solar in metallicity. Moreover, they found good agreement between the photometric [Fe/H] values derived by Nemec et al. (2011) for 19 non-Blahko stars and the spectroscopic [Fe/H] values.

\section{What's next?}

{\bf Kepler's Future} 
Unfortunately, since May 2013 Kepler's pointing accuracy has been compromised by the breakdown of a second of its four reaction wheels, which keep it aligned. In August 2013 the Kepler team ended its attempts to restore the spacecraft to full working order. It is now being considered what kind of research can be done with Kepler in its current condition. The KASC asteroseismic community has contributed several white papers to guide the decision.

In the meantime, we have plenty of existing Kepler data to examine, in addition to the ground-based multicolour photometry (Young-Beom et al. 2013) and spectroscopy (Nemec et al. 2013). With only a fraction of the Kepler RR Lyrae data analyzed in detail so far, we can expect to learn more in the years to come.

{\bf Upcoming Space Missions}  
The NASA mission TESS (Transiting Exoplanet Survey Satellite), a worthy successor to the Kepler mission was selected for launch in 2017. It can be expected that this all-sky planet-hunting survey will yield ultra-precise light curve of many more (Blazhko) RR Lyrae stars as a "by-product".

The European Space Agency's global space astrometry mission GAIA is scheduled to be launched later this year (2013). It will map the position, motion, luminosity, temperature and composition of probably tens of thousands of RR Lyrae stars.  These (versatile) data will undoubtedly contribute to another great leap in our understanding of these stars, even if GAIA's duty cycle (on average 70 measurements over a five-year period) may not allow us to identify all the Blazhko stars with accurate periods.
If the ESA mission candidate PLATO2.0 gets selected, it will also have an important asteroseismology component. 
%The Canadian space mission $MOST$ is still going strong after 10 years in space, and has done important observations of the RRd star AQ Leo (Gruberbauer et al. 2007).

\begin{flushleft}
\begin{deluxetable*}{llll}
\tabletypesize{\normalsize}
\tablecaption{Published Research on Kepler RR Lyrae stars and spin-offs}
\tablewidth{0pt}
\tablehead{ \colhead{Topic}   & \colhead{Authors} &
       \colhead{No. of stars} & \colhead{Remark\footnotemark[a]} }

\startdata
First results (period doubling) &  Kolenberg et al. (2010)  &  2 & A \\
Period doubling in Kepler data & Szabo et al. (2010) & 3+4 & A, M \\
Flavors of variability & Benk\H{o} et al. (2010) & 29 & A \\
RR Lyr long-cadence data & Kolenberg et al. (2011) & 1 & A \\
Testing Stothers model on RR Lyr & Smolec et al. (2011) & 1 & A, M\\
"Crazy star" V445 Lyr & Guggenberger et al. (2012) & 1 & A \\
Non-modulated RR Lyrae stars & Nemec et al. (2011) & 19 & A \\
RRc multiperiodicity & Moskalik et al. (2012) & 4 & A \\
Nonlinear asteroseismology for RR Lyr & Molnar et al. (2012) & 1 &  A, M \\
$[Fe/H]$, $v_{\rm rad}$, etc. from spectra   & Nemec et al. (2013) & 41 & A, G \\
RR Lyr in short cadence & Stellingwerf et al. \footnotemark[b] & 1 & A\\
Ground-based multicolor data & Jeon et al. (2013) & 41 & G \\
Additional RR Lyrae stars?  & Kinemuchi\footnotemark[b]  & TBD & A, G \\
Mathematical description of modulation & Benk\H{o} et al. (2010) & & M \\
Radial resonance Blazhko model & Buchler \& Koll\'ath (2011) & & M \\
Shock Blazhko model & Gillet (2013) &  & M \\
FFI photometry paper & Kinemuchi et al. (in prep.) & TBD & A \\
\enddata
\footnotetext[a]{A: Analysis, M: Modelling,  G: ground-based data, S: Spin-off}
\footnotetext[b]{These proceedings}
\end{deluxetable*}
\end{flushleft}

\pagebreak
\section{Acknowledgements}

I cordially thank the organizers of this conference, Horace's former PhD students, for making this a very enjoyable and special meeting, and a special thanks to Karen for enabling me to attend.  
Horace, thank you for your great contributions to variable star research and RR Lyrae studies in particular, and for making this fascinating research field accessible to so many!
I am grateful for the support from Marie Curie Fellowship 255267 SAS-RRL within the 7th European Community Framework Program.

\end{document}